\documentclass[useAMS,usenatbib]{mn2e}
\usepackage{graphicx}
\usepackage{amssymb}
\usepackage{lscape}
\usepackage{ulem}
\usepackage{txfonts}

\def\kms{km~s$^{-1}$}
\def\cm{cm$^{-2}$}
\def\lya{Ly$\alpha$}
\def\nhi{$N$(H\,{\sc i})}
\def\hi{H\,{\sc i}}
\def\si2{Si\,{\sc ii}}
\def\mg2{Mg\,{\sc ii}}
\def\fe2{Fe\,{\sc ii}}
\def\al2{Al\,{\sc ii}}
\def\zn2{Zn\,{\sc ii}}
\def\c2s{C\,{\sc ii}$^{\star}$}

\def\dind{$D$-index}

\title[The kinematic signature of DLAs]{The kinematic signature of damped Lyman alpha systems: Using the D-index to screen for high column density
HI absorbers\thanks{Based on observations made 
with ESO Telescopes at the Paranal Observatories under programmes
078.A-0003(A) and 080.A-0014(B).}}

\author[Ellison et al.] {Sara L. Ellison$^1$, Michael T. Murphy$^2$,  
Miroslava Dessauges-Zavadsky$^3$.\\
$^1$Department of Physics and Astronomy, University of Victoria, Victoria, B.C., V8P 1A1, Canada\\
$^2$Centre for Astrophysics and Supercomputing, Swinburne University of
Technology, Mail H39, PO Box 218, Victoria 3122, Australia\\
$^3$Geneva Observatory, University of Geneva, 51 Ch. des Maillettes,
1290 Sauverny, Switzerland}

\begin{document}

\maketitle

\begin{abstract}
Using a sample of 21 damped Lyman alpha systems (DLAs) 
and 35 sub-DLAs, we evaluate the \dind\
$\equiv \frac{EW (\AA)}{\Delta v (km/s)} \times 1000$ from
high resolution spectra of the \mg2\ $\lambda$ 2796 profile.
This sample represents an increase in sub-DLA \dind\ statistics
by a factor of four over the sample used by Ellison (2006).
We investigate various techniques to define the velocity spread ($\Delta v$)
of the \mg2\ line to determine an optimal \dind\ for the identification
of DLAs.  The success rate of DLA identification is 50 -- 55\%,
depending on the velocity limits used, improving
by a few percent when the column density of \fe2\ is included in
the \dind\ calculation.  We recommend the set of 
parameters that is judged to be most robust, have a combination
of high
DLA identification rate (57\%) and low DLA miss rate (6\%) and
 most cleanly separate
the DLAs and sub-DLAs (Kolmogorov-Smirnov probability 0.5\%).  
These statistics demonstrate that the \dind\ is 
the most efficient technique for selecting low redshift DLA candidates:
65\% more efficient than selecting DLAs based on the equivalent
widths of \mg2\ and \fe2\ alone. 
We also investigate the effect of resolution on determining the \nhi\
of sub-DLAs.
We convolve echelle spectra of sub-DLA \lya\ profiles with Gaussians
typical of the spectral resolution of instruments on the Hubble
Space Telescope and compare the best fit \nhi\ values at both
resolutions.  We find that the fitted \hi\ column density is systematically
over-estimated by $\sim 0.1$ dex in the moderate resolution spectra compared
to the best fits to the original echelle spectra.  
This offset is due to blending
of nearby \lya\ clouds that are included in the damping wing fit
at low resolution. 

\end{abstract}


\section{Introduction}

The near-UV is a critical wavelength regime for quasar absorption
line spectroscopy.  Whilst blue-sensitive ground-based instruments
such as UVES at the VLT have opened the door for such efforts as
measuring the molecular content of DLAs (Ledoux et al. 2003;
Noterdaeme et al. 2008) and the study of the low redshift \lya\
forest (e.g. Kim et al. 2002), studying neutral hydrogen at $z < 1.6$
requires a space telescope.  Moreover, due to the declining
number density of galactic scale absorbers, such as damped Lyman
alpha systems (DLAs), at low redshifts, blind searches for 
these galaxies place
unrealistic demands on space resources.  Therefore, whilst the number of known
DLAs at $z>1.7$ is now in excess of 1000, thanks to trawling
large ground-based optical surveys
(http://www.ucolick.org/$\sim$xavier/SDSSDLA/index.html), the head-count
of low redshift DLAs is only around 5\% of the high $z$ tally
(e.g. Rao, Turnshek \& Nestor 2006).  Characterising the low-to-intermediate
redshift DLA population is important not just for the large fraction
of cosmic time that it represents, but also because the detection
of galactic counterparts for the absorbers is largely only feasible
at $z_{\rm abs}<1$.

In order to circumvent the high cost of a blind survey for low $z$
DLAs, the practice in recent years has been to pre-select DLA
candidates based on the detection of strong metal lines in ground-based
spectra (e.g. Rao \& Turnshek 2000).  Although there is no tight 
correlation between the \nhi\ and \mg2\ equivalent width (EW), 
there is a statistical correlation
for large samples, e.g. Bouch\'e (2008) and M\'enard \& Chelouche (2008).
In the most recent survey of low $z$ DLAs selected by metal lines,
Rao et al. (2006) found that 36\% of absorbers 
with rest frame EWs of  \mg2\ $\lambda$ 2796 and \fe2\ $\lambda$ 2600 
$>$ 0.5 \AA\ 
were DLAs.  Including the additional criterion that the EW of
 Mg\,{\sc i} $\lambda > 0.1$ \AA, increases
success rate for identifying DLAs to 42\%.  The remaining
absorbers were found to have 18.0 $<$ log \nhi\ $<$ 20.3 (Table 1
of Rao et al. 2006), thus spanning the range from
Lyman limit systems (LLS) to sub-DLAs.  Whereas DLAs have neutral hydrogen
column densities \nhi\ $\ge 2 \times 10^{20}$\cm, sub-DLAs are 
usually defined (e.g. Peroux et al. 2003a) to have 19.0 $<$ log \nhi\ 
$<$ 20.3.  The upper bound of this classification corresponds to the
transition to classical DLAs, whilst the lower limit is driven
by the clarity of the \lya\ damping wing necessary to produce
a reliable fit.  The LLS and sub-DLAs are usually excluded from
the statistical calculation of quantities such as $\Omega_{\rm gas}$
(the mass density of neutral gas).  The contribution of
sub-DLAs to $\Omega_{\rm gas}$, and even the validity of including
the sub-DLAs in the census for neutral gas are still hotly debated 
topics (e.g. Peroux et al. 2003b; Prochaska, Herbert-Fort \& Wolfe 2005).
Nonetheless, the sub-DLAs are emerging as an interesting field
of research in their own right, and for statistical purposes it is
often desirable to separate the DLAs and sub-DLAs (e.g.
Turnshek \& Rao 2002).  For these reasons, it is desirable to develop
an empirical tool that allows an observer to pre-select candidate
absorption systems that are most suitable for their purpose.
Therefore, whilst the \mg2\ + \fe2\ EW selection
has certainly been the key to identifying the vast majority of low
redshift DLAs, a more robust separation of DLAs and sub-DLAs from
ground-based spectra would be welcome.

Ellison (2006) proposed a new way to screen \mg2\ absorbers
as potential DLAs, defining the \dind\ as the ratio between
\mg2\ EW and velocity width (see Section \ref{d_sec}).
For a sample of 27 absorbers,  the success rate of
the \dind\ in selecting DLAs was found to be 
more than a factor of two improvement
over the use of EW alone.  However, only eight of the 27 absorbers
were sub-DLAs, and yet these lower column density absorbers
are more abundant than their high \nhi\ cousins (e.g. Peroux
et al. 2003b).  In this work, we re-visit the assessment of the
\dind\ as a tool for the pre-selection of DLAs based on
high resolution spectroscopy of \mg2\ absorbers with an enlarged
sample of 56 absorbers.  

\section{Sample selection}\label{sample_sec}

\begin{center}
\begin{table*}
\caption{New QSO observations}
\begin{tabular}{lllllll}
\hline
QSO & $z_{\rm em}$ &  V & Instrument & Observation date & Exp. time (s) & S/N per pixel \\  
\hline 
 Q0009-016 &  1.998 & 17.6 & UVES  & Nov. 2006 & 6000 & 25 \\
Q0021+0043 &  1.245 & 17.7 & UVES  & Nov. 2006 & 9000 & 20 \\
Q0157-0048 &  1.548 & 17.9 & UVES  & Oct. 2006 & 6000 & 20 \\
 Q0352-275 &  2.823 & 17.9 & UVES  & Oct. \& Nov. 2006 & 10240 & 35 \\
 Q0424-131 &  2.166 & 17.5 & UVES  & Oct. \& Nov. 2006 & 6000 & 30 \\
Q1009-0026 &  1.244 & 17.4 & UVES  & Jan. 2007 & 6000 & 30 \\
Q1028-0100 &  1.531 & 18.2 & UVES  & Feb. 2007 & 10240 & 20 \\
Q1054-0020 &  1.021 & 18.3 & UVES  & Jan. \& Feb. 2007 & 10240 & 35 \\
 Q1327-206 &  1.165 & 17.0 & UVES  & Feb. 2008 & 4460 & 25 \\
Q1525+0026 &  0.801 & 17.0 & HIRES & Jul. 2006 & 1200 & 10 \\
 Q2048+196 &  2.367 & 18.5 & HIRES & Jul. 2006 & 3655 & 20 \\
Q2328+0022 &  1.308 & 17.9 & HIRES & Jul. 2006 & 2500 & 10 \\
Q2352-0028 &  1.628 & 18.2 & HIRES & Jul. 2006 & 5400 & 20 \\
 \hline 
\end{tabular}\label{new_obs}
\end{table*}
\end{center}

\begin{center}
\begin{table*}
\caption{Full absorber sample for D-index calculations}
\begin{tabular}{lcccll}
\hline
QSO & $z_{\rm abs}$ &  log \nhi\  & N(\fe2)  & \nhi\ reference &  \fe2\ reference \\
\hline 
 Q0002+051 & 0.851 & 19.08 $\pm$ 0.04  &  13.89$\pm$0.04 &  Rao, Turnshek \& Nestor (2006)       & Murphy, unpublished		   \\
 Q0009-016 & 1.386 & 20.26 $\pm$ 0.02  &  14.32$\pm$0.04 &  Rao, Turnshek \& Nestor (2006)       & Dessauges-Zavadsky et al. in prep.  \\
Q0021+0043 & 0.520 & 19.54 $\pm$ 0.03  &  13.17$\pm$0.04 &  Rao, Turnshek \& Nestor (2006)       & Dessauges-Zavadsky et al. in prep.  \\
Q0021+0043 & 0.940 & 19.38 $\pm$ 0.15  &  14.62$\pm$0.04 &  Rao, Turnshek \& Nestor (2006)       & Dessauges-Zavadsky et al. in prep.  \\
 Q0058+019 & 0.613 & 20.08 $\pm$ 0.20  &  15.24$\pm$0.06 &  Pettini et al. (2000)                & Pettini et al. (2000)    	   \\
 Q0100+130 & 2.309 & 21.37 $\pm$ 0.08  &  15.09$\pm$0.01 &  Dessauges-Zavadsky et al. (2004)     & Dessauges-Zavadsky et al. (2004)    \\
 Q0117+213 & 0.576 & 19.15 $\pm$ 0.07  &   ...           &  Rao, Turnshek \& Nestor (2006)       & ...				   \\
Q0157-0048 & 1.416 & 19.90 $\pm$ 0.07  &  14.57$\pm$0.03 &  Rao, Turnshek \& Nestor (2006)       & Dessauges-Zavadsky et al. in prep.  \\
 Q0216+080 & 1.769 & 20.00 $\pm$ 0.20  &  14.53$\pm$0.09 &  Lu et al. (1996)                     & Dessauges-Zavadsky et al. in prep.  \\
 Q0352-275 & 1.405 & 20.18 $\pm$ 0.18  &  15.10$\pm$0.03 &  Rao, Turnshek \& Nestor (2006)       & Dessauges-Zavadsky et al. in prep.  \\
 Q0424-131 & 1.408 & 19.04 $\pm$ 0.04  &  13.45$\pm$0.02 &  Rao, Turnshek \& Nestor (2006)       & Dessauges-Zavadsky et al. in prep.  \\
 Q0454-220 & 0.474 & 19.45 $\pm$ 0.03  &   ...           &  Rao, Turnshek \& Nestor (2006)       & ...				   \\
Q0512-333A & 0.931 & 20.49 $\pm$ 0.08  &  14.47$\pm$0.06 &  Lopez et al. (2005)                  & Lopez et al. (2005)		   \\
Q0512-333B & 0.931 & 20.47 $\pm$ 0.08  &$>$14.65         &  Lopez et al. (2005)                  & Lopez et al. (2005)		   \\
 Q0823-223 & 0.911 & 19.04 $\pm$ 0.04  &  13.57$\pm$0.04 &  Rao, Turnshek \& Nestor (2006)       & Meiring et al. (2007)		   \\
  Q0827+24 & 0.525 & 20.30 $\pm$ 0.05  &  14.59$\pm$0.02 &  Rao \& Turnshek (2000)                & Meiring et al. (2006)		   \\
 Q0841+129 & 2.375 & 20.99 $\pm$ 0.08  &  14.76$\pm$0.01 &  Dessauges-Zavadsky et al. (2006 )    & Dessauges-Zavadsky et al. (2006)    \\
Q0957+561A & 1.391 & 20.30 $\pm$ 0.10  &  14.46$\pm$0.05 &  Churchill et al. (2003a)             & Churchill et al. (2003a) 	   	   \\
Q0957+561B & 1.391 & 19.90 $\pm$ 0.10  &  14.34$\pm$0.05 &  Churchill et al. (2003a)             & Churchill et al. (2003a) 	   	   \\
Q1009-0026 & 0.840 & 20.20 $\pm$ 0.06  &  14.48$\pm$0.02 &  Rao, Turnshek \& Nestor (2006)       & Dessauges-Zavadsky et al. in prep.  \\
Q1009-0026 & 0.880 & 19.48 $\pm$ 0.08  &  14.37$\pm$0.07 &  Rao, Turnshek \& Nestor (2006)       & Dessauges-Zavadsky et al. in prep.  \\
Q1010-0047 & 1.327 & 19.81 $\pm$ 0.07  &  14.53$\pm$0.03 &  Rao, Turnshek \& Nestor (2006)       & Meiring et al. (2007)		   \\
Q1028-0100 & 0.709 & 20.04 $\pm$ 0.07  &  15.10$\pm$0.03 &  Rao, Turnshek \& Nestor (2006)       & Dessauges-Zavadsky et al. in prep.  \\
Q1054-0020 & 0.951 & 19.28 $\pm$ 0.02  &  13.71$\pm$0.01 &  Rao, Turnshek \& Nestor (2006)       & Dessauges-Zavadsky et al. in prep.  \\
 Q1101-264 & 1.838 & 19.50 $\pm$ 0.05  &  13.51$\pm$0.02 &  Dessauges-Zavadsky et al. (2003)     & Dessauges-Zavadsky et al. (2003)    \\
Q1104-180A & 1.662 & 20.85 $\pm$ 0.01  &  14.77$\pm$0.02 &  Lopez et al (1999)                   & Lopez et al (1999) 		   \\
 Q1122-168 & 0.682 & 20.45 $\pm$ 0.05  &  14.55$\pm$0.01 &  Ledoux, Bergeron \& Petitjean (2002)  & Ledoux, Bergeron \& Petitjean (2002) \\
 Q1151+068 & 1.774 & 21.30 $\pm$ 0.08  &   ...           &  Dessauges-Zavadsky, unpublished      & ...				   \\
 Q1157+014 & 1.944 & 21.60 $\pm$ 0.10  &  15.46$\pm$0.02 &  Dessauges-Zavadsky et al. (2006)     & Dessauges-Zavadsky et al. (2006)    \\
 Q1206+459 & 0.928 & 19.04 $\pm$ 0.04  &  12.95$\pm$0.02 &  Rao, Turnshek \& Nestor (2006)       & Murphy, unpublished		   \\
 Q1210+173 & 1.892 & 20.63 $\pm$ 0.08  &  15.01$\pm$0.03 &  Dessauges-Zavadsky et al. (2006)     & Dessauges-Zavadsky et al. (2006)    \\
 Q1213-002 & 1.554 & 19.56 $\pm$ 0.02  &  14.44$\pm$0.01 &  Rao, Turnshek \& Nestor (2006)       & Murphy, unpublished		   \\
 Q1223+175 & 2.466 & 21.44 $\pm$ 0.08  &  15.16$\pm$0.02 &  Dessauges-Zavadsky, unpublished      & Prochaska et al. (2001)		   \\
Q1224+0037 & 1.235 & 20.88 $\pm$ 0.06  &$>$15.11         &  Rao, Turnshek \& Nestor (2006)       & Meiring et al. (2007)		   \\
Q1224+0037 & 1.267 & 20.00 $\pm$ 0.08  &  14.54$\pm$0.09 &  Rao, Turnshek \& Nestor (2006)       & Meiring et al. (2007)		   \\
 Q1247+267 & 1.223 & 19.88 $\pm$ 0.10  &  13.97$\pm$0.04 &  Pettini et al. (1999)                & Pettini et al. (1999)		   \\
 Q1327-206 & 0.853 & 19.40 $\pm$ 0.02  &  13.90$\pm$0.04 &  Rao, Turnshek \& Nestor (2006)       & Dessauges-Zavadsky et al. in prep.  \\
 Q1331+170 & 1.776 & 21.14 $\pm$ 0.08  &  14.63$\pm$0.03 &  Dessauges-Zavadsky et al. (2004)     & Dessauges-Zavadsky et al. (2004)    \\
 Q1351+318 & 1.149 & 20.23 $\pm$ 0.10  &  14.74$\pm$0.07 &  Pettini et al. (1999)                & Pettini et al. (1999)		   \\
 Q1451+123 & 2.255 & 20.30 $\pm$ 0.15  &  14.33$\pm$0.07 &  Dessauges-Zavadsky et al. (2003)     & Dessauges-Zavadsky et al. (2003)    \\
Q1525+0026 & 0.567 & 19.78 $\pm$ 0.08  &  14.19$\pm$0.06 &  Rao, Turnshek \& Nestor (2006)       & Dessauges-Zavadsky et al. in prep.  \\
 Q1622+238 & 0.656 & 20.40 $\pm$ 0.10  &   ...           &  Rao \& Turnshek (2000)                & ...				   \\
 Q1622+238 & 0.891 & 19.23 $\pm$ 0.03  &   ...           &  Rao \& Turnshek (2000)                & ...				   \\
 Q1629+120 & 0.900 & 19.70 $\pm$ 0.04  &   ...           &  Rao, Turnshek \& Nestor (2006)       & ...				   \\
 Q2048+196 & 1.116 & 19.26 $\pm$ 0.08  &$>$15.22         &  Rao, Turnshek \& Nestor (2006)       & Dessauges-Zavadsky et al. in prep.  \\
 Q2128-123 & 0.430 & 19.37 $\pm$ 0.08  &$>$14.08         &  Ledoux, Bergeron \& Petitjean (2002)  & Ledoux, Bergeron \& Petitjean (2002) \\
 Q2206-199 & 1.920 & 20.44 $\pm$ 0.08  &  15.30$\pm$0.02 &  Pettini et al. (2002)                & Pettini et al. (2002) 		   \\
 Q2230+023 & 1.859 & 20.00 $\pm$ 0.10  &  ...            &  Dessauges-Zavadsky et al. (2006)     & ...    \\
 Q2231-001 & 2.066 & 20.53 $\pm$ 0.08  &  14.83$\pm$0.03 &  Dessauges-Zavadsky et al. (2004)     & Dessauges-Zavadsky et al. (2004)    \\
Q2328+0022 & 0.652 & 20.32 $\pm$ 0.07  &  14.84$\pm$0.01 &  Rao, Turnshek \& Nestor (2006)       & Dessauges-Zavadsky et al. in prep.  \\
Q2331+0038 & 1.141 & 20.00 $\pm$ 0.05  &  14.38$\pm$0.03 &  Rao, Turnshek \& Nestor (2006)       & Meiring et al. (2007)		   \\
 Q2343+125 & 2.431 & 20.35 $\pm$ 0.05  &  14.66$\pm$0.03 &  Dessauges-Zavadsky et al. (2004)     & Dessauges-Zavadsky et al. (2004)    \\
 Q2348-144 & 2.279 & 20.59 $\pm$ 0.08  &  13.84$\pm$0.05 &  Dessauges-Zavadsky et al. (2004)     & Dessauges-Zavadsky et al. (2004)    \\
Q2352-0028 & 0.873 & 19.18 $\pm$ 0.10  &  13.47$\pm$0.06 &  Rao, Turnshek \& Nestor (2006)       & Dessauges-Zavadsky et al. in prep.  \\
Q2352-0028 & 1.032 & 19.81 $\pm$ 0.14  &  14.96$\pm$0.04 &  Rao, Turnshek \& Nestor (2006)       & Dessauges-Zavadsky et al. in prep.  \\
Q2352-0028 & 1.246 & 19.60 $\pm$ 0.30  &  14.28$\pm$0.03 &  Rao, Turnshek \& Nestor (2006)       & Dessauges-Zavadsky et al. in prep.  \\    
\hline 
\end{tabular}\label{qso_table}
\end{table*}
\end{center}

We selected sub-DLAs from the compilation
of Rao et al. (2006) for follow-up spectroscopy at high resolution,  
obtaining new echelle spectra of 13 QSOs with 17 absorbers along their
lines of sight.
Nine of these QSOs were observed with the UV and Visual Echelle Spectrograph
(UVES) at the Very Large Telescope (VLT)
and four with High Resolution Echelle Spectrograph (HIRES) 
at the Keck telescope. The observations and
data reduction are described in detail in Dessauges-Zavadsky
et al. (in preparation).  In Table \ref{new_obs} we summarise 
the details of these new data; emission redshifts and $V$-band magnitudes
are taken from Rao et al. (2006).  The new data were reduced using the publically
available pipelines UVES$\_$popler (e.g. Zych et al. 2008, see also
http://astronomy.swin.edu.au/$\sim$mmurphy/UVES$\_$popler) and xidl HIRES redux (e.g.
Bernstein et al. in preparation,
see also http://www.ucolick.org/$\sim$xavier/HIRedux/index.html). We have also added
to the archival sample, thanks to the donation of spectra that
have appeared in Meiring et al. (2006, 2007)
and Churchill et al. (1999, 2003b).  Finally, one additional spectrum (Q0216+080)
has been obtained from the UVES archive and re-reduced by
us (Dessauges-Zavadsky et al. in prep.).  Our final sample
consists of 56 absorbers, of which 21 are DLAs and 35 are
sub-DLAs.   We have therefore doubled the total sample size
of absorbers since Ellison (2006) and increased the
number of sub-DLAs by more than a factor of four.  
In Table
\ref{qso_table} we list the full absorber sample, absorber
redshifts, \hi\ and \fe2\ column densities and references for
these quantities\footnote{In cases where errors on N(\fe2) are not
available in the literature , we assign a value of 0.05 dex.}.  
Out of the 56 absorbers in our sample, 49
have available N(\fe2) column densities, of which 19 are DLAs and 30 are sub-DLAs.

\section{Results}

\subsection{Sub-DLAs at moderate resolution}\label{res_sec}

Most of the sub-DLAs in our sample have been drawn from the
surveys of Rao \& Turnshek (2000) and Rao et al. (2006), where
the \nhi\ column densities have been determined from moderate
resolution HST spectra. Although the damping wings of sub-DLAs
should theoretically be present in data with resolutions below that of echelle
spectrographs (typically FWHM $\sim$ 6 \kms, or 0.1 \AA), there
are a number of practical problems which may affect the \lya\ fit at
all resolutions, 
including blending of nearby \lya\ forest lines, the accuracy of the 
continuum fit and correct sky subtraction.  Therefore, although
random errors can be quoted for the \nhi\ determinations (typically
0.05 -- 0.10 dex), it is important to also consider any systematic
effects.  This is particularly relevent for a technique that aims
to distinguish DLAs and sub-DLAs, since we must be confident
that the absorbers are correctly classified.  Indeed, Meiring et al.
(2008) have previously suggested that low resolution spectra
might lead to over-estimates of \nhi\ when only single absorption
components are fitted.  Here, we quantitatively investigate this
possibility and assess its impact on the current study.

\begin{figure}
\centerline{\rotatebox{0}{\resizebox{8cm}{!}
{\includegraphics{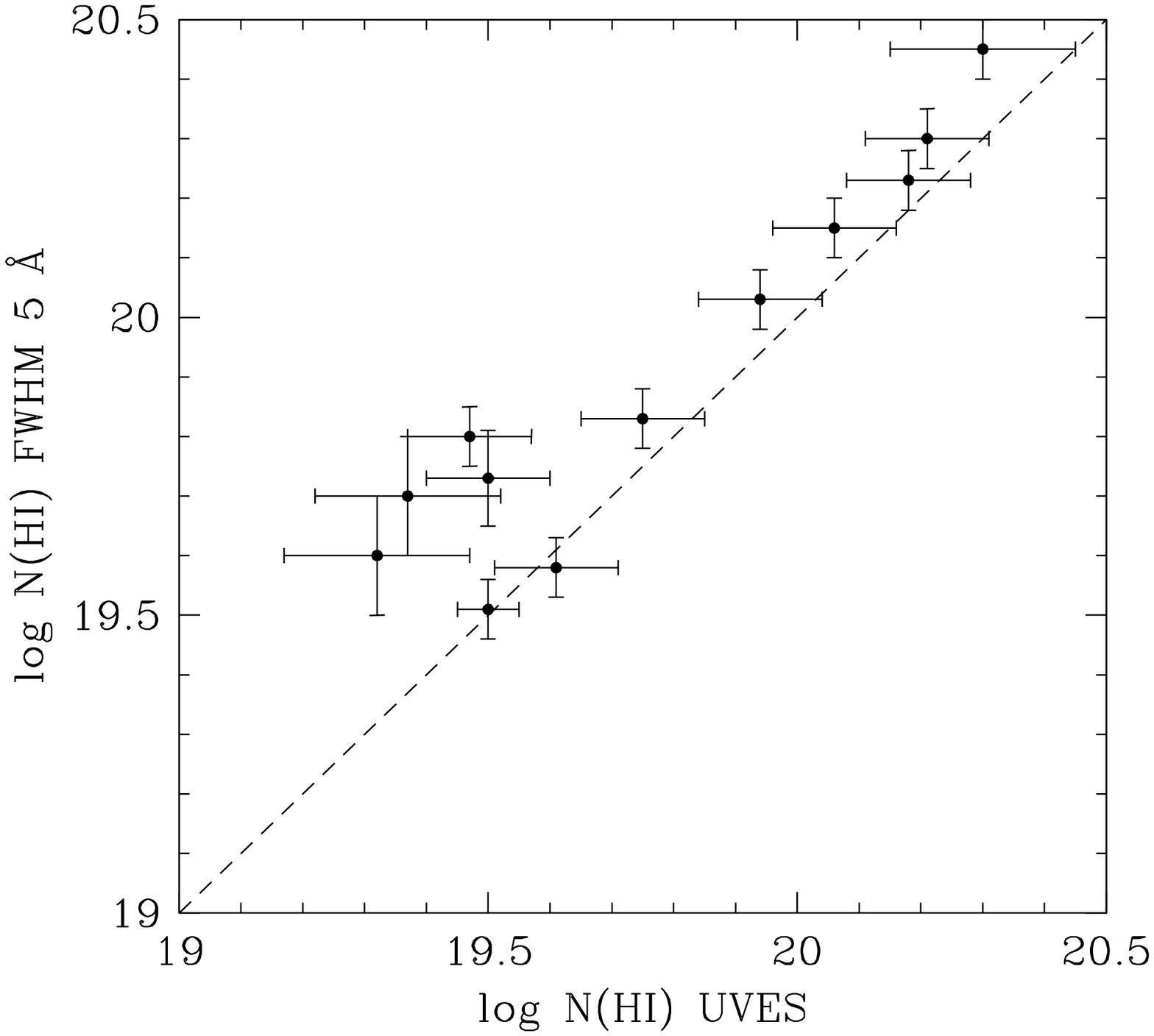}}}}
\caption{Comparison of the best fit \nhi\ column densities
from high resolution UVES data and VPFITs to the UVES
data after convolution with a gaussian of FWHM = 5 \AA.
The column densities are given in Table \ref{conv_tab}\label{nhi_comp_fig} }
\end{figure}

The HST spectra used by Rao \& Turnshek (2000) and Rao et al. (2006)
have been compiled from various surveys and archival data from the
Faint Object Spectrograph (FOS) and
Space Telescope Imaging Spectrograph (STIS; both MAMA and CCD observations).  
Although the data vary in
characteristics, the typical FWHM resolution $\sim$ 5 \AA.
In order to investigate the presence of systematic uncertainties
in the \nhi\ determinations of sub-DLAs, we have taken the UVES
spectra of the 12 sub-DLAs published by Dessauges-Zavadsky et al.
(2003) and convolved them with a gaussian profile of FWHM=5\AA\
to simulate the effect of HST spectral resolution.  The process
of convolution effectively smooths the noise out of the UVES data.
However, we do not re-insert any noise characteristics, so that
our fitting tests are driven by the effects of resolution, not
noise.  We use VPFIT\footnote{http://www.ast.cam.ac.uk/$\sim$rfc/vpfit.html.} 
to determine
the \hi\ column density of absorbers in the convolved spectra.  In Table
\ref{conv_tab} we give the original UVES best fit \nhi\ values
from Dessauges-Zavadsky et al. (2003) and our VPFIT values.
The comparison is shown visually in Figure \ref{nhi_comp_fig}.

The main result of this test is that we find a \textit{systematic}
offset between \nhi\ determined from the high resolution 
UVES data and the spectra that have been convolved to mimic
HST resolution.  The low resolution fits yield \nhi\ values
that are typically higher than the UVES-determined values by 0.1
dex, although the discrepancy can be as large as 0.3 dex.
The magnitude of the discrepancy appears to be driven by
the amount of local \lya\ absorption.  At high resolution,
it is relatively straightforward to distinguish low column
density \lya\ forest clouds from the main sub-DLA absorption.
However, at low resolution, these components become indistinguishable
and the fit is driven to higher values of \nhi\ to account for the
increased equivalent width.

\begin{figure*}
\centerline{\rotatebox{270}{\resizebox{14cm}{!}
{\includegraphics{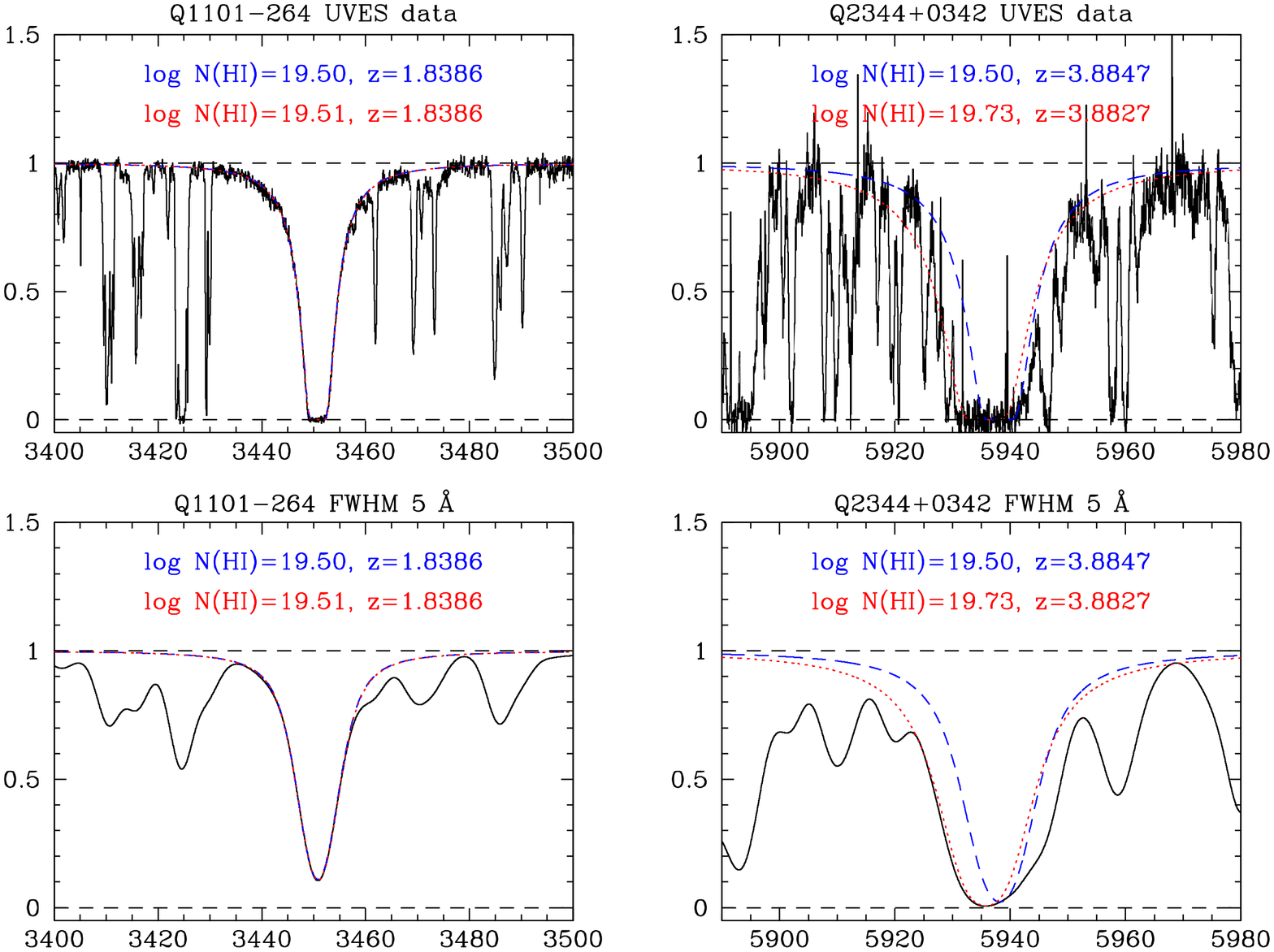}}}}
\caption{Two examples of fits to high resolution (top panels) and
FWHM=5 \AA\ (bottom panels) for the sub-DLAs towards Q1101$-$264
(left panels) and Q2344$+$0342 (right panels).  In all
panels the red dotted line shows the damped profile of the best
fit to the low resolution data (see red panel labels) and
the blue dashed line shows the fit by Dessauges-Zavadsky et al.
(2003) to the UVES data (blue panel labels). The profiles have
been convolved to the resolution of the data in each panel.
This Figure demonstrates that when the sub-DLA is isolated,
the \nhi\ determination is excellent, even at moderate resolution.
However, when \lya\ forest blending becomes significant, there
is a tendency to over-estimate the \nhi\ from low
resolution data.  \label{res_test_fig} }
\end{figure*}

This effect is demonstrated in Figure \ref{res_test_fig},
where we show two extreme cases, Q1101$-$264 and Q2344+0342.  The former
of these sub-DLAs appears isolated from \lya\ forest blending
in the UVES spectra and has a very well defined continuum
and consequently an excellent \hi\ fit.  Conversely,
Q2344+0342 is highly blended and Dessauges-Zavadsky et al (2003)
found it necessary to simultaneously fit a second absorption
system in the blue wing of the sub-DLA in order to obtain
a satisfactory fit.  In Figure \ref{res_test_fig} we show
both the original UVES data (top panels) and convolved
data (lower panels) for the two QSOs.  In each panel, the
blue dashed lines show the profiles (at the appropriate resolution
for the data shown) for the \nhi\ values derived from the UVES
data and the red dotted lines show the values determined
from VPFIT to the convolved data.  Therefore, although the
\nhi\ values for the top and bottom panels are identical for
a given sub-DLA, the resolution of the profile is matched to the
effective instrumental resolution of the data shown.  For convenience, the
\nhi\ values are given (with the appropriate colour code) at the top
of each panel.  Since redshift is also a free parameter
in VPFIT, we also show its best-fit value, as well as the value
obtained from the UVES data (taken from Dessauges-Zavadsky et al.
2003).  In the case of Q1101$-$264, it can be seen that the
VPFIT \nhi\ column density and redshift are in excellent agreement
with the UVES values.  The same is not true of the sub-DLA towards
Q2344+0342.  From the bottom-right panel, we see that the
UVES \nhi\ value (blue dashed line) is a poor fit to the low
resolution data.  The VPFIT solution (red dotted line)
is driven to fit the blue
wing, which, in reality, is severely blended with additional
\lya\ absorption.  This also causes the redshift of the two solutions
to differ by 120 \kms.  In the top-right panel, it can be
seen how the VPFIT low resolution fit is a poor representation
of the high resolution data.  

\begin{center}
\begin{table}
\caption{Resolution test: \nhi\ fitting results}
\begin{tabular}{lccc}
\hline
QSO & $z_{\rm abs}$ & UVES \nhi & FWHM=5\AA\ \nhi \\
\hline
Q1101$-$264 & 1.838  & 19.50$\pm$0.05 & 19.51$\pm$0.05 \\ 
Q1223+175 & 2.466  & 19.32$\pm$0.15 & 19.60$\pm$0.10 \\ 
Q1409+095 & 2.668   & 19.75$\pm$0.10 & 19.83$\pm$0.05 \\ 
Q1444+014 & 2.087  & 20.18$\pm$0.10 & 20.23$\pm$0.05 \\ 
Q1451+123 & 2.255  & 20.30$\pm$0.15 & 20.45$\pm$0.05 \\ 
Q1511+090 & 2.088  & 19.47$\pm$0.10 & 19.80$\pm$0.05 \\ 
Q2059$-$360 & 2.507  & 20.21$\pm$0.10 & 20.30$\pm$0.05 \\ 
Q2116$-$358 & 1.996  & 20.06$\pm$0.10 & 20.15$\pm$0.05 \\ 
Q2155+1358 & 3.142 & 19.94$\pm$0.10 & 20.03$\pm$0.05 \\ 
~~~~~~~~~~ & 3.565 & 19.37$\pm$0.15 & 19.70$\pm$0.10 \\ 
~~~~~~~~~~ & 4.212 & 19.61$\pm$0.10 & 19.58$\pm$0.05 \\ 
Q2344+0342 & 3.882  & 19.50$\pm$0.10 & 19.73$\pm$0.08 \\ 
\hline 
\end{tabular}\label{conv_tab}
\end{table}
\end{center}

Although we have demonstrated that there is a tendency to
over-estimate \nhi\ from low resolution spectra, the effect,
in general, is relatively small, usually $\sim$ 0.1 dex (see
Table \ref{conv_tab}).  Abundances of sub-DLAs may therefore
have been previously under-estimated by 0.1 -- 0.3 dex, and
there are also implications for the number densities of
absorbers with modest HI column densities.
At low redshift, this effect may be mediated somewhat by the
lower line number density as the forest thins.  On the other
hand, it has been shown that sub-DLAs are often found near
(in velocity space) to other strong absorbers (e.g. Ellison \& Lopez 2001;
Peroux et al. 2003a; Ellison et al. 2007).  Statistics that
rely on weighted, or total \nhi\ values should not be
greatly affected by this systematic error, since the fractional
contribution of blended \lya\ forest absorption will have
only a minor impact on the high column density DLAs that
dominate such quantities.  

Finally, we note that the systematic effect investigated
here does not include errors in the continuum fit.
The error in continuum fitting is difficult to quantify for
absorbers in general, given that different authors use very different
techniques, and that these techniques themselves are often
adapted to the resolution and S/N of the data.  Errors in
the continuum are often estimated to add a further 0.1 dex
uncertainty to the determination of \nhi.

\subsection{The D-index}\label{d_sec}

Ellison (2006) defined the D-index to be

\begin{equation}\label{d_eqn}
D = \frac{EW (\AA)}{\Delta v (km/s)} \times 1000,
\end{equation}

\noindent where the EW is that of the \mg2\ $\lambda$ 2796 line and $\Delta v$
is the velocity spread of the same line.
The limits of $\Delta v$ were determined by excluding `detached'
absorption components where the continuum is recovered,
and only include the complex with the
largest EW.  The limits of $\lambda_{\rm min, max}$ (the minimum
and maximum wavelengths over which the EW is calculated) were set
where the absorption becomes significant at the 3 $\sigma$ level below
the continuum.

With a larger sample, we can experiment further with the
definition of $D$, in order to optimize the distinction between
DLAs and sub-DLAs.  We begin by experimenting with different
ways of defining $\lambda_{\rm min, max}$ in the \mg2\ $\lambda$
2796 \AA\ line.  We do not appeal to additional lines at this
stage, since we are aiming to develop a method that can be
used with the limited spectral information that is
available for low redshift absorbers.

\begin{figure}
\centerline{\rotatebox{270}{\resizebox{6cm}{!}
{\includegraphics{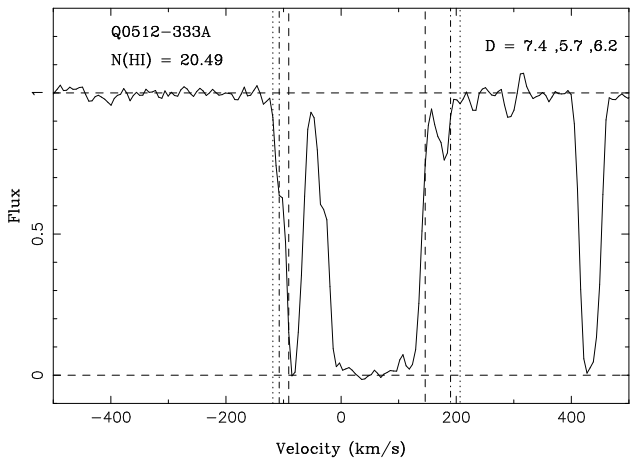}}}}
\caption{The \mg2\ $\lambda$ 2796 transition for one of the
absorbers in our sample.  The vertical lines show the minimum
and maximum velocities used to calculate the D-index 
for the three different possibilities investigated in this paper: 
90\% of the equivalent width (dashed), 90\% of
the velocity width (dot-dashed) and 3 $\sigma$ absorption significance 
(dotted). The respective D-indices calculated from these 3
techniques are listed in the top right of the panel.  \mg2\
spectra are given for the full sample in the online material
that accompanies this article.
\label{lam_lim}}\end{figure}

The values of EW and $\Delta v$ that are input into
Eqn \ref{d_eqn} are governed by the choice of 
$\lambda_{\rm min, max}$.  We have investigated the following possibilities:
(i)  $\lambda_{\rm min, max}$ defined by 3 $\sigma$ absorption
(described above and in Ellison 2006);
(ii) $\lambda_{\rm min, max}$ defined by the central 90\% of the \mg2\ line
EW.  This is analogous to the technique
used on unsaturated lines to determine the velocity
width of an absorber based on optical depth (e.g. Prochaska \&
Wolfe 1997; Ledoux et al. 2006);  (iii) $\lambda_{\rm min, max}$ 
defined by the central 90\% of the velocity spread of the \mg2\ line. 
As a further experiment, we apply these three techniques
using the full spread of all \mg2\ components in our
calculation of the \dind, in addition to the original
definition of Ellison (2006) which uses only the main absorption complex.
The D-index calculated over the central absorption excludes
outlying components once the continuum has been recovered,
whereas `full' refers to all absorption components in a
given \mg2\ system.
In Figure \ref{lam_lim} we show how these different approaches
yield different values for $\lambda_{\rm min, max}$ in one
of the absorbers in our sample.  This Figure is reproduced for all of our 
absorber sample in the online material that accompanies this article.
We also provide in the online material a table listing all of the
EWs and errors for the 3 approaches described above for each absorber;
Table \ref{ew_table} gives the first four lines of the online version
as an example.

\begin{center}
\begin{table*}
\caption{Equivalent widths and velocity spreads for lambda limits
discussed in the text (see online material for the full table). 
}
\begin{tabular}{lccccccc}
\hline
 QSO    &     log N(HI) &   EW$_{90\%EW}$(\AA)  & $\Delta v_{90\%EW}$(\kms) &  EW$_{90\%vel}$(\AA)  &   $\Delta v_{90\%vel}$ (\kms) & EW$_{3\sigma}$ (\AA) &  $\Delta v_{3\sigma}$ (\kms) \\ \hline
 Q0823$-$223 & 19.04$\pm$0.04  &  0.85$\pm$0.03  &  186.38 & 0.93$\pm$0.04 &  231.30  & 0.93$\pm$0.04   &  206.59\\
 Q1206$+$459 & 19.04$\pm$0.04  &  0.56$\pm$0.02  &  106.86  &  0.63$\pm$0.03 &  142.48  &    0.63$\pm$0.03    &     115.76\\
 Q0424$-$131 & 19.04$\pm$0.04 &   0.25$\pm$0.01  & 37.53 &  0.28$\pm$0.02 & 65.04  &   0.26$\pm$0.02    &  40.03 \\
 Q0002$+$051 & 19.08$\pm$0.04  &  0.62$\pm$0.03    &  95.03 &  0.66$\pm$0.04   &  129.80   & 0.63$\pm$0.02  &  85.76\\
\hline 
\end{tabular}\label{ew_table}
\end{table*}
\end{center}

\begin{figure}
\centerline{\rotatebox{270}{\resizebox{6cm}{!}
{\includegraphics{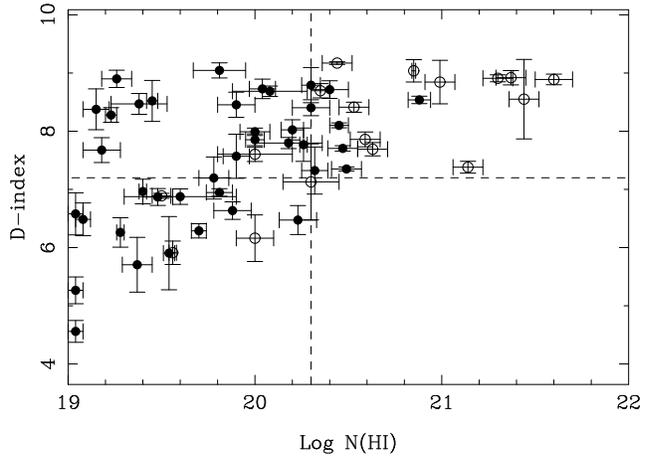}}}}
\caption{The \dind\ versus \nhi\ for the central range of \mg2\ 
absorption $\lambda_{\rm min, max}$ as defined by
the 90\% EW range.  The horizontal dashed line shows the
optimal \dind\ as given in Table \ref{d_tab} and the
vetrical dashed line demarcates DLAs and sub-DLAs.  The
open points have $z_{\rm abs} > 1.5$ and filled points have
$z_{\rm abs} \le 1.5$, see the discussion in Section \ref{use_sec}.
\label{d_vs_nhi}}
\end{figure}

In Table \ref{d_tab}
we give the success statistics and optimal $D$ value for each
technique\footnote{The D-index statistics which include N(\fe2)
are described later in this section.}.  
The success rate is defined as the fraction of systems
with $D \ge D_{\rm optimal}$ that are actually DLAs.  The DLA miss
rate is the fraction of DLAs with $D < D_{\rm optimal}$.
We also tabulate the Kolmogorov-Smirnov (KS) 
probability that the $D$-indices of
sub-DLAs and DLAs for a given technique are drawn from the same
population.  Table \ref{d_tab} shows that all of the techniques 
yield more or less consistent
results, with success rates $\sim$ 50 -- 55 \% and KS probabilities
on the order of a few percent.  As an example, we show the
distribution of $D$ versus \nhi\ in Figure \ref{d_vs_nhi}
when the 90\% EW limits are applied to the central absorption
complex.

\begin{center}
\begin{table*}
\caption{D-index statistics for $\lambda_{\rm min, max}$ tests}
\begin{tabular}{llcccc}
\hline
Absorption range & $\lambda$ limits & Optimal $D$ & DLA success rate (\%) & DLA miss rate (\%) & KS probability (\%)\\ 
\hline
Central & 3 $\sigma$ & 6.7 & 48.6 & 19.0 & 9.1 \\
Central &  90\% velocity & 5.5 & 54.3 & 9.5 & 0.1 \\
Central & 90\% EW & 7.2 & 54.1 & 4.8 & 0.4 \\
Full & 3 $\sigma$ & 6.3 & 51.5 & 19.0  & 5.3 \\
Full & 90\% velocity & 5.2 & 53.1 & 19.0 & 2.4 \\
Full & 90\% EW & 7.0 & 56.7 & 19.0 & 1.3 \\
Central + N(\fe2) & 3 $\sigma$ & 6.3 & 48.4 & 11.8 & 20.6 \\
Central + N(\fe2) &  90\% velocity & 5.7 & 57.7 & 11.8 & 0.1 \\
\textbf{Central + N(\fe2)} & \textbf{90\% EW} & \textbf{7.0} & \textbf{57.1} & \textbf{5.9} & \textbf{0.5} \\
Full + N(\fe2) & 3 $\sigma$ & 5.2 & 55.6 & 16.7  & 1.3 \\
Full + N(\fe2) & 90\% velocity & 5.1 & 60.0 & 16.7 & 0.6 \\
Full + N(\fe2) & 90\% EW & 5.6 & 53.6 & 16.7 & 1.3 \\
\hline 
\end{tabular}\label{d_tab}
\end{table*}
\end{center}

We next consider whether including information from
other metal lines may improve the use of $D$ to distinguish
DLAs and sub-DLAs.   It has been
shown (e.g. Meiring et al. 2008; Peroux et al. 2008) 
that sub-DLAs have a tendency to be
more metal-rich than DLAs.  Indeed, there is a
lack of high \nhi\, high metallicity systems that
empirically manifests itself as an apparent anti-correlation
between \nhi\ and [Zn/H] (Boisse et al. 1998;
Prantzos \& Boissier 2000).  Although it has been
argued that this is due to dust bias, this interpretation
is not supported by extinction measurements of DLAs
(Ellison, Hall \& Lira 2005).  Simply put, the
observed reddening values derived from DLA samples
(Murphy \& Liske 2004; Ellison et al. 2005; Vladilo,
Prochaska \& Wolfe 2008) are much lower than would
be required to impose a `dust filter'.  Moreover,
it has been argued (Dessauges-Zavadsky et al. 2003) 
that the higher metallicities
in sub-DLAs are not a result of ionization corrections.
As we have shown above, any systematic error in the
\nhi\ of low redshift sub-DLAs determined from low
resolution spectra is both small, and in the wrong sense
to explain the different metallicity distributions.
In the absence of any experimental reason (such as dust bias
or ionization correction) behind the elevated metallicities
in sub-DLAs, we explore whether information on the metallicity
of an absorber can be combined with the kinematic
description encapsulated in Eqn \ref{d_eqn} to
improve the selection success of the \dind.

\begin{figure}
\centerline{\rotatebox{270}{\resizebox{6cm}{!}
{\includegraphics{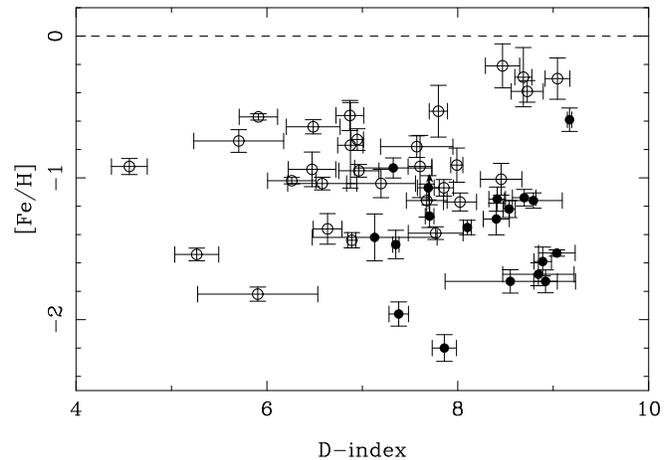}}}}
\caption{The \dind\ versus [Fe/H] for the central range of \mg2\ 
absorption $\lambda_{\rm min, max}$ as defined by
the 90\% EW range. Solid points denote DLAs, open
points denote sub-DLAs. The horizontal dashed line
represents solar iron abundance, for which we adopt the
value log(N(Fe)/\hi)$_{\odot}$ = $-4.55$ (Asplund et al. 2005).
\label{d_vs_feh}}
\end{figure}

\begin{figure}
\centerline{\rotatebox{0}{\resizebox{8cm}{!}
{\includegraphics{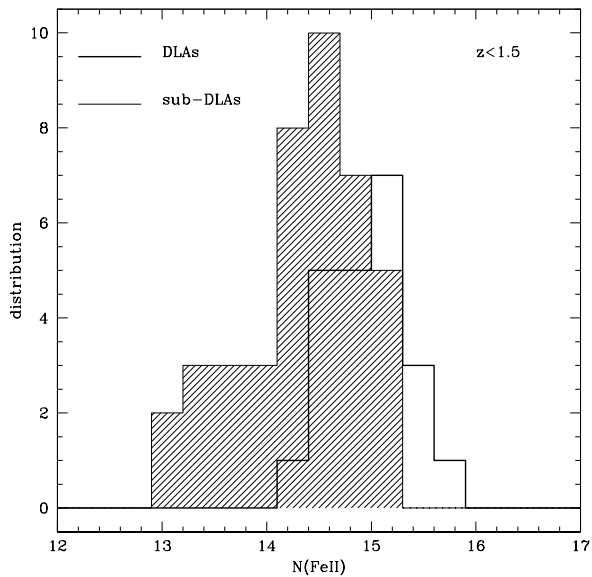}}}}
\caption{Distribution of \fe2\ column densities for DLAs (open histogram,
bold outline) and sub-DLAs (shaded histogram).\label{fe_dist} }
\end{figure}

In Figure \ref{d_vs_feh} we show the distribution
of [Fe/H] versus \dind\  (as defined by
the 90\% EW range for the central absorption component)
for the 49 absorbers for
which N(\fe2) measurements are available in the
literature.  It can be seen that in the regime of
high \dind, the DLAs have preferentially lower metallicities
than the sub-DLAs.  For $D\ge 7$, the mean [Fe/H] is
$-0.67$ for sub-DLAs and $-1.43$ for DLAs. These
mean values are calculated by treating the lower
limits of the three absorbers with fully saturated
\fe2\ profiles as detections.  The difference between
the mean metallicities could therefore be even larger. 
Although the DLAs in our sample have
a higher mean redshift than the sub-DLAs (see next
section) the evolution of DLA metallicity is very mild.
Kulkarni et al. (2005) show that the DLA metallicity
changes by less than 0.3 dex from $z \sim 0.5$ to $z \sim 2.5$.
Of course, we can
not use the actual metallicity to fine-tune the definition
of the \dind, since this requires
a measurement of \nhi, the very quantity we are hoping to
select for.  We therefore consider the use of N(\fe2).
In Figure \ref{fe_dist} we show histograms
of log N(\fe2) for the compilation of DLAs and sub-DLAs in
Dessauges-Zavadsky et al. (in prep.)
at $z_{\rm abs} < 1.5$ (the redshift regime for
which we are trying to develop a selection technique).
The column density of \fe2\ is a quantity that can
be measured with relative ease from almost any high
resolution spectrum suitable for \dind\ determination.
There are almost a dozen different \fe2\ transitions with
a large dynamic range in $f$-value at wavelengths
not far from \mg2.  
From Figure \ref{fe_dist} we can see that
the DLAs clearly have systematically higher N(\fe2)
than the sub-DLAs.  The mean log N(\fe2) (and standard deviation)
is 15.15$\pm$0.08 for DLAs and 14.64$\pm$0.07 for sub-DLAs.
Although the element zinc is
usually considered the best indicator of actual metallicity, we
emphasize here that we are simply using iron from the
empirically motivated viewpoint that its column density
is systematically different in DLAs and sub-DLAs. 

We factor N(\fe2) into our calculation of the \dind\ by multiplying
the value derived from Eqn \ref{d_eqn} by log N(\fe2)
and dividing by 15 (to re-scale the D-index):

\begin{equation}\label{dfe_eqn}
D = \frac{EW (\AA)}{\Delta v (km/s)} \times \frac{log N(FeII)}{15} \times 1000
\end{equation}

\begin{figure}
\centerline{\rotatebox{270}{\resizebox{6cm}{!}
{\includegraphics{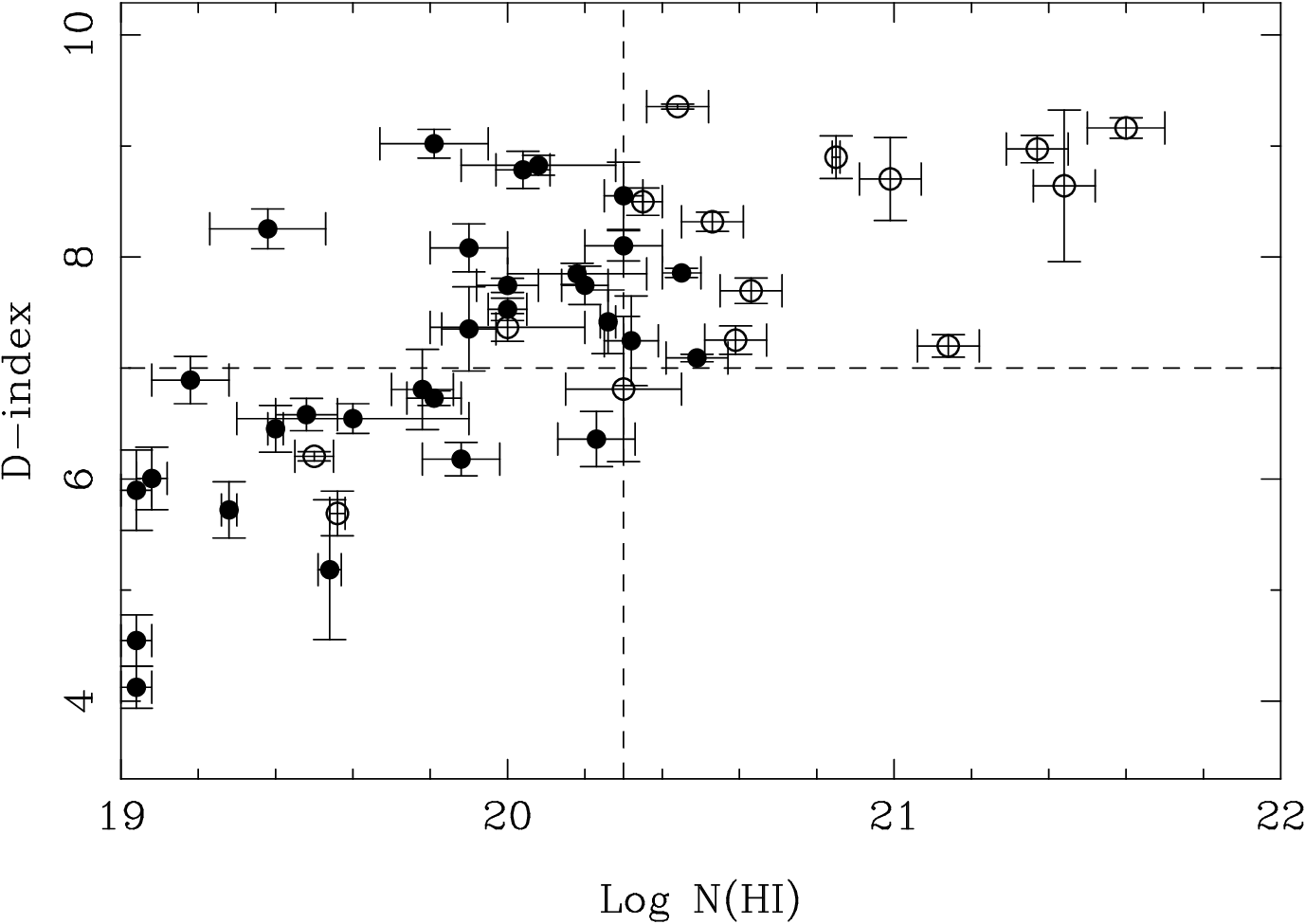}}}}
\caption{The \dind\ versus \nhi\ for the central range of \mg2\ 
absorption $\lambda_{\rm min, max}$ as defined by
the 90\% EW range. The \dind\ has been calculated from Eqn \ref{dfe_eqn}.  
These data points refer to the
row in bold in Table \ref{d_tab}.  The horizontal dashed line shows the
optimal \dind\ as given in Table \ref{d_tab} and the
vetrical dashed line demarcates DLAs and sub-DLAs.  The
open points have $z_{\rm abs} > 1.5$ and filled points have
$z_{\rm abs} \le 1.5$, see the discussion in Section \ref{use_sec}.
\label{d_vs_nhi_fe}}
\end{figure}

In Table \ref{d_tab} we give the success and miss rates for
this modified \dind\ definition, again calculating its value for
the six different combinations investigated above and excluding 
systems which only have N(\fe2) limits.  Figure \ref{d_vs_nhi_fe}
shows the distribution of \dind\ versus \nhi\ for
our sample using the definition of D given in Eqn \ref{dfe_eqn}.
Although the success rates generally improve slightly (by a few
percent) when N(\fe2) is included, this is not a significant improvement
given the modest-sized sample.  The
close consistency of success rates with/without N(\fe2) included 
in the calculation is probably due to the small relative difference 
(in the log) between iron column densities. It is also worth noting
that in Figure \ref{d_vs_feh} there is apparently no correlation
between \dind\ and [Fe/H] in DLAs, indicating that the \dind\
does not preferentially select high or low metallicity DLAs.  On the
other hand, 4/5 sub-DLAs with $D>8$ have [Fe/H] $> - 0.5$, the highest
values in our sample.  Therefore, using low values of $D$ to select
sub-DLAs may miss the highest metallicity absorbers.

\section{Summary and discusssion}

We have presented a sample of 56 \mg2\ absorbers with known
\nhi\ column densities and investigated techniques to screen
for DLAs.  The \dind\ combines measures of both \mg2\ EW
and velocity spread, and we have also investigated incorporating
information on the \fe2\ column density.  The results are
robust to the various parameterizations that we use, with a
typical DLA yield of $\sim$ 55\%, compared to $\sim$ 35\%
when \mg2\ EW alone is used (Rao et al. 2006).
The relative insensitivity to the precise definition of the \dind\ 
is reassuring, in the sense that
it demonstrates that its success is unlikely to be a fluke of our 
dataset or choice of parameterization.

\subsection{Using the \dind}\label{use_sec}

Given the results in Table \ref{d_tab}, what is the `best' definition
of $D$?   Given the sample size (and hence, rounding errors in the
success and miss rates), simply selecting the technique with the
best statistics is not necessarily the best choice.  Instead, we
consider which technique might be the most robust against issues
such as noise, resolution and absorber environment.  Any of the
methods that use the `full' velocity range will be sensitive to outlying
components that may have nothing to do with the main absorber, e.g.
be associated with a satellite galaxy.  As discussed in Section
\ref{res_sec}, sub-DLAs in particular are known to often have
companion absorbers.  S/N is a
consideration for techniques that use a 3 $\sigma$
cut-off, either over the full absorption range, or just the central
complex.  However, Ellison (2006) showed that S/N only significantly
degrades the efficiency of \dind\ selection in the central absorption
component  at S/N $\ll 10$.  
The disadvantage of using the 90\% velocity range is that
it is sensitive to small differences (between spectra) of 
resolution. As Ellison (2006) showed, this is also the case for a
3 $\sigma$ definition.
Taking these issues into account, and then re-visiting the
statistics of Table \ref{d_tab}, we suggest that the
90\% EW over the central absorption range (i.e excluding outlying,
low EW clouds), should be the
most robust to these various troubles and yields a high
success rate, with only one DLA missed.  The optimal values of
$D$ are very similar for the central/90\% EW technique whether or
not N(\fe2) is included -- 7.0 and 7.2 respectively.  The KS test
probability is slightly better when \fe2\ is included.

One potential bias in the evaluation of the \dind\ in the
current sample is that the redshift distributions of the
DLAs and sub-DLAs are not the same.  The mean DLA redshift
is $z_{\rm abs} = 1.63$, compared with 1.06 for the sub-DLAs,
with very few of the latter above $z_{\rm abs} = 1.5$.
This potential bias can be seen in Figures \ref{d_vs_nhi}
and \ref{d_vs_nhi_fe} where the open points show absorbers at $z_{\rm abs}
> 1.5$.  Confining our analysis to absorbers only in this low redshift
range includes only 5 DLAs with \fe2\ detections, 
insufficient for robust statistical
analysis.  
However, Mshar et al. (2007) have shown that
neither the EW distribution, nor the velocity spread of strong
\mg2\ systems evolves with redshift.  These authors do, however,
find differences in the number and distribution of `subsystems'
between their low and high redshift sample.  Whilst the low redshift
absorbers tend to comprise a single strong absorption component
with one satellite subsystem, there are more subsystems in
high redshift absorbers, over the same velocity range.  The
\dind\ is sensitive to such differences.  However, the redshift
differences in kinematic structure reported by Mshar et al. (2007)
and described above
can not explain the `success' of the \dind\ in distinguishing
DLAs from sub-DLAs.  This is because the redshift inequality
in our sample is such that we have more DLAs at high $z$,
where the tendency towards more complex kinematic structure
would decrease the \dind.  However, DLAs are defined by high values
of $D$, i.e. with few kinematically extended subsystems.  Further
evidence against a redshift bias in our sample comes from Ledoux et al.
(2006) who find that, in DLAs, the median velocity width of unsaturated
metal lines increases with decreasing redshift.  This redshift dependence
again works in the opposite sense to a redshift bias that would artificially
introduce a \dind\ dependence in our work.  Therefore,
although it is certainly desirable to repeat the \dind\
tests for a sample of sub-DLAs and DLAs that are matched in
redshift, there is no empirical evidence to suggest that
redshift evolution can explain the difference in $D$.

In summary, we recommend that the \dind\ is calculated using the
90\% EW range over the central absorption complex, incorporating
N(\fe2), if available, see Equation \ref{dfe_eqn}.  
In Figure \ref{d_vs_nhi_fe} we show the distribution
of \nhi\ versus \dind\ for this choice of parameters.  This
Figure is directly comparable to Figure \ref{d_vs_nhi} which
uses the same definition of the \dind, but does not include
N(\fe2) in the calculation.
Based on our recommended definition, the KS probability that
the $D$-indices of sub-DLAs and DLAs are drawn from the same population
is 0.5\%.  In our sample of 45 absorbers with accurate \fe2\ column densities, 
57$^{+18.3}_{-14.0}$\% (16/28) of systems with $D >$ 7.0,
are DLAs where the error bars are 1 $\sigma$ values taken from Gehrels
(1986). Only 6\% (1/17) of the DLAs in the sample of 45 have $D\le7.0$.

\subsection{Interpretation of the \dind}

Ellison (2006) suggested that the \dind\ may be driven in part
by the QSO-galaxy impact parameter.  For a galaxy composed of
numerous \mg2\ `clouds', whose size is of the order of a few
kpc (Ellison et al. 2004b), the absorption may appear more patchy
when the sightline intersects at large impact parameter where the
filling factor of these clouds is low.
A sightline passing close to the galaxy's centre will intercept
a higher density of absorbing clouds, perhaps with a smaller
radial velocity distribution.  This picture is supported by
empirical evidence that the EW of \mg2\ absorbers does
correlate with impact parameter (Churchill et al. 2000; Chen \&
Tinker 2008) and sub-DLAs do have more distinct absorbing
components than DLAs (Churchill et al. 2000).

Another possible source of distinction between the kinematic
structure of DLAs and sub-DLAs is the presence of galactic outflows.
The connection between outflows and a subset of strong \mg2\ 
absorbers has been discussed
in the literature for several years (e.g. Bond et al. 2001;
Ellison, Mallen-Ornelas \& Sawicki 2003), although it remains
contentious as a general description (see discussions
in Bouch\'e et al. 2006 and  Chen \& Tinker 2008).
Nonetheless, galactic outflows are apparently common in high
redshift star-forming galaxies (e.g. Shapley et al 2003; Weiner et al. 2008)
and a growing body of empirical evidence is being
published which supports the connection of strong \mg2\ absorbers
with winds or outflows.  For example, Zibetti et al. (2007)
find that the mean impact parameter of \mg2\ absorbers in
the SDSS is about 50 kpc from a L$^{\star}$ galaxy.  One
explanation for this observation is that the absorption detected
in QSO spectra corresponds to material that is in a large extended
region, possibly associated with winds. M\'enard
\& Chelouche (2008) also argue for the connection of \mg2\ absorbers
to massive galaxies based on their gas-to-dust ratios, which
are an order of magnitude higher than values for DLAs (Murphy
\& Liske 2004; Ellison et al. 2005; Vladilo et al. 2006).  
Murphy et al. (2007) observed a significant correlation between
\mg2\ EW and metallicity and concluded that `[there is] a strong link
between absorber metallicity and the mechanism for producing and
dispersing the velocity components'.  Finally, Kacprzak et al. (2007)
have shown that there is a correlation between the \mg2\ EW and the
degree of asymetry in the host galaxy.
However, it does not necessarily flow that \textit{all} strong \mg2\ 
absorbers are connected with outflows.
   Both Bouch\'e (2008) and M\'enard \& Chelouche (2008)
have recently suggested that, when plotted in the
\nhi-\mg2\ EW plane, there are two separate absorber populations
distinguished by metallicity.  This effect is also seen in Figure
\ref{d_vs_feh} where the high \dind\ absorbers are fairly cleanly
separated into high metallicity sub-DLAs and low metallicity
DLAs. Bouch\'e (2008) proposes
that the more metal-rich, lower \nhi\ absorbers at a given
\mg2\ EW are associated with galactic outflows.  In turn,
this may lead to kinematics that are more patchy in velocity
space (e.g. Ellison et al. 2003) and contribute to the mechanism
behind the success of the \dind.

\subsection{Applications of the \dind}

The \dind\ requires moderately high resolution spectra to 
yield accurate pre-selection.  Therefore, whilst EW-only DLA
pre-selection can exploit the truly enormous datasets of, 
for example, the SDSS (e.g. Rao \& Turnshek 2000; Rao et al.
2006), what is the niche for the \dind?    Individual
groups have already used moderately large datasets for \mg2\
searches at high resolution.       The two most recent
surveys for strong \mg2\ absorbers in high resolution spectra are
Mshar et al. (2007) (56 absorbers with $\lambda$ 2796 EW $>$ 0.3 \AA)
and Prochter et al. (2006a) (21 absorbers with \mg2\ $\lambda$ 2796 
EW $>$ 1 \AA).  Given that the number density of \mg2\ 
$\lambda$ 2796 EW $>$ 0.6 \AA\ absorbers is a factor 2
larger than at 1 \AA, these two samples would already yield over
50 absorbers suitable for \dind\ screening (not accounting
for an overlap between the samples).  However, these samples
are small compared to the full archival power that could
be applied.  After a decade of high
resolution spectroscopy with HIRES on Keck and UVES at the VLT,
and a somewhat shorter, but extremely productive history with
ESI, we estimate that 500 -- 700 QSOs have been observed at 
moderate--high resolution (e.g. Prochaska et al. 2007b).  
If we assume an average emission
redshift $z_{\rm em}$=2.5 for 500 QSOs, and maximum wavelength 
coverage of 8500
\AA, this yields a redshift path of almost 760.  The number
density of strong \mg2\ absorbers has been robustly determined
from SDSS spectra (e.g. Nestor et al. 2005; Prochter et al. 2006a).  
For example, Nestor et al. (2005) find that the number
density of \mg2\ $\lambda$ 2796 EW $>$ 0.6 \AA\ absorbers
(a typical EW suitable for DLA pre-selection) at $z \sim 1$
is $\sim 0.5$.  This value is unlikely to be affected due to
magnitude bias (Ellison et al. 2004a)
and can therefore also be applied to high resolution
spectra.  The total number\footnote{We note that
the sample used for the study in this paper
does not have much overlap with this archival path length.}
of such absorbers that we
could therefore expect to find in existing moderate--high
resolution spectra is therefore approaching 400.  Even
though follow-up observations with space telescopes are likely to focus
on the brightest in this sample, the typical magnitude limit
for echelle spectroscopy with an 8-m ground-based telescope
is close to the limit that will be feasible for the Cosmic
Origins Spectrograph (COS) and the renovated STIS . 

The \dind\ may also be useful for assessing the nature of absorption
where follow-up observations of \lya\ are not even possible.  For example, in
the spectra of GRBs.  As observatories (and observers) fine-tune
their rapid follow-up techniques, there is a growing database
of bright GRBs that have been observed promptly at high resolution 
(e.g. Vreeswijk et al. 2007;
Ellison et al. 2007; Prochaska et al. 2007a).  \mg2\ absorbers detected
in these spectra that do not have \lya\ covered, have no chance
of susbsequent follow-up due to the rapid fading of
the optical afterglow, but their likelihood of being DLAs can still
be assessed, (e.g. Ellison et al.  2006; Prochaska et al. 2007a).
The application to GRB spectra is particularly interesting
given the possibility of deep searches for the absorbing
galaxy after the optical afterglow has faded.  Even at low
redshifts, the study of DLA host galaxies has been challenging
with QSO impact parameters of 1--2 arcseconds (e.g. Chen \& Lanzetta 2003
and references therein).  
Identifying probable DLAs in GRB spectra opens the door to host galaxy
searches to unprecedented depths.

Finally, in light of the recent puzzling result by Prochter et al. (2006b)
that intervening \mg2\ absorbers are more numerous towards GRB 
sightlines than QSO sightlines, it would also be interesting
to compare the distribution of $D$-indices of these two populations.

\section*{Acknowledgments} 
SLE was funded by an NSERC Discovery grant.   MTM thanks the
Australian Research Council for a QEII Research Fellowship (DP0877998).
We are indebted to
Chris Churchill and Joseph Meiring for providing spectra that expanded our
archival sample.
Joe Hennawi provided valuable help in obtaining and reducing the HIRES data.
This research has made use of the NASA/IPAC Extragalactic 
Database (NED) which is operated by the Jet Propulsion Laboratory, 
California Institute of Technology, under contract with the National 
Aeronautics and Space Administration.


\begin{thebibliography}{}
\small
\itemindent -0.48cm

\bibitem[Asplund et al. (2005)]{asp05}
        Asplund, M., Grevesse, N., Sauval, A. J., 2005
	ASP Conference Series, 336, 25

\bibitem[Boisse et al (1998)]{boi98}
	Boisse, P., Le Brun, V., Bergeron, J., Deharveng, J.-M.,
	1998, A\&A, 333, 841

\bibitem[Bond et al (2001)]{bond01b}
        Bond, N., Churchill, C., Charlton, J., Vogt, S., 2001, ApJ, 562, 641

\bibitem[Bouch\'e et al. 2006]{bou06}
        Bouch\'e, N., Murphy, M. T., Peroux, C., Csabai, I., Wild, V.,
	2006, MNRAS, 371, 495

\bibitem[Bouch\'e (2008)]{bou08}
        Bouch\'e, N., 2008, arXiv:0803.3944v2

\bibitem[Chen \& Lanzetta (2003)]{cl03}
        Chen, H.-W., Lanzetta, K., 2003, ApJ, 597, 706

\bibitem[Chen \& Tinker (2008)]{ct08}
        Chen, H.-W., Tinker, J. L., 2008, ApJ, submitted, arXiv:0801.2169

\bibitem[Churchill et al (1999)]{cwc99}
        Churchill, C. W., Rigby, J R., Charlton, J. C., Vogt, S. S.,
	1999, ApJS, 120, 51

\bibitem[Churchill et al (2003a)]{cwc03a}
        Churchill, C. W., Mellon, R. R., Charlton, J. C., Vogt, S.,
 	2003a, ApJ, 593, 203 %Q0957+561

\bibitem[Churchill et al (2003b)]{cwc03b}
        Churchill, C. W.,  Vogt, S., Charlton, J. C., 
 	2003b, AJ, 125, 98 %vpfits to mg2

\bibitem[Dessauges-Zavadsky et al. (2003)]{dz03}
         Dessauges-Zavadsky, M., P\'eroux, C., Kim, T.-S., D'Odorico, S., 
	 McMahon, R. G., 2003, MNRAS, 345, 447

\bibitem[Dessauges-Zavadsky et al (2004)]{mikr04}
        Dessauges-Zavadsky, M., Calura, F., Prochaska, J. X., 
	D'Odorico, S., Matteucci, F., 2004, A\&A, 416, 79

\bibitem[Dessauges-Zavadsky et al. (2006)]{dz06}
         Dessauges-Zavadsky, M., Prochaska, J. X., D'Odorico, S., 
	 Calura, F., Matteucci, F., 2006, A\&A, 445, 93

\bibitem[Ellison \& Lopez (2001)]{el01}
        Ellison, S. L., \& Lopez, S., 2001, A\&A, 380, 117

\bibitem[Ellison, Mallen-Ornelas \& Sawicki (2003)]{q1331}
        Ellison, S. L., Mallen-Ornelas, G., \& Sawicki, M., 2003,
	ApJ,589, 709

\bibitem[Ellison et al. (2004)]{corals3}
        Ellison, S. L., Churchill, C. W., Rix, S. A., Pettini, M.,
	2004a, ApJ, 615, 118

\bibitem[Ellison et al. (2004)]{apm04}
        Ellison, S. L., Ibata, R., Pettini, M., Lewis, G. F., Aracil, B., 
	Petitjean, P., Srianand, R., 2004b, A\&A, 414, 79

\bibitem[Ellison, Hall \& Lira (2005)]{coralsbmk}
        Ellison, S. L., Hall, P. B., Lira, P., 2005, AJ, 130, 1345 

\bibitem[Ellison (2006)]{dindex}
        Ellison, S. L., 2006, MNRAS, 368, 335

\bibitem[Ellison et al. (2006)]
         Ellison, S. L., et al. 2006, MNRAS, 372, L38

\bibitem[Ellison et al. (2007)]{pair}
        Ellison, S. L., Hennawi, J. F., Martin, C. L., Sommer-Larsen, J.,
	2007, MNRAS, 378, 801

\bibitem[Gehrels, N. (1986)]{geh86}
        Gehrels, N., 1986, ApJ, 303, 336

\bibitem[Kacprzak et al. (2007)]{gk07}
         Kacprzak, G. G.,  Churchill, W. C., Steidel, C. C., Murphy, M.
	 T., Evans, J. L., 2007, ApJ, 662, 909

\bibitem[Kim et al. (2002)]{kim02}
         Kim, T.-S., Carswell, R. F., Cristiani, S., D'Odorico, S., 
	 Giallongo, E., 2002, MNRAS, 335, 555

\bibitem[Kulkarni et al. (2005)]{kul05}
        Kulkarni, V., Fall, S. M., Lauroesch, J. T., York, D. G., Welty,
	D. E., Khare, P., Truran, J., W., 2005, ApJ, 618, 68

\bibitem[Ledoux, Bergeron \& Petitjean (2002)]{lbp02}
	Ledoux, C., Bergeron J., \& Petitjean, P., 2002,
	A\&A, 305, 802

\bibitem[Ledoux et al. (2003)]{led03}
        Ledoux, C., Petitjean, P., Srianand, R., 2003, MNRAS,
	346, 209

\bibitem[Ledoux et al. (2006)]{led06}
	Ledoux, C., Petitjean, P., Fynbo, J. P. U., Moller, P., Srianand, R.,
	2006, A\&A, 457, 71

\bibitem[Lopez et al 1999]{seb99}
	Lopez, S., Reimers, D., Rauch, M., Sargent, W., Smette, A.,
	1999, ApJ, 513, 598

\bibitem[Lopez et al. (2005)]{seb05}
        Lopez, S., Reimers, D., Gregg, M. D., Wisotzki, L., Wucknitz, O., 
	Guzman, A., 2005, ApJ, 626, 767 

\bibitem[Meiring et al. (2006)]{mei06}
        Meiring, J., Kulkarni, V. P., Khare, P., Bechtold, J.,  
	York, D. G., Cui, J., Lauroesch, J. T., Crotts, A. P. S.Nakamura, O.,
	2006, MNRAS, 370, 43

\bibitem[Meiring et al. (2007)]{mei07}
        Meiring, J. D., Lauroesch, J. T., Kulkarni, V. P., Peroux, C., 
	Khare, P., York, D. G., Crotts, A. P. S., 2007, MNRAS, 376, 557

\bibitem[Meiring et al. (2008)]{mei08}
        Meiring, J. D., Kulkarni, V. P., Lauroesch, J. T., Peroux, C., 
	Khare, P., York, D. G., Crotts, A. P. S., 2008, MNRAS, 384, 1015

\bibitem[M\'enard \& Chelouche (2008)]{mc08}
         M\'enard, B., \& Chelouche, D., 2008, MNRAS, submitted, arXiv:0803.0745

\bibitem[Murphy \& Liske (2004)]{ml04}
        Murphy, M. T., \& Liske, J., 2004, MNRAS, 345, L31

\bibitem[Murphy et al. (2007)]{mim07}
         Murphy, M. T., Curran, S. J., Webb, J. K., Menager, H., Zych, B. J.,
	 2007, MNRAS, 376, 673

\bibitem[Nestor, Turnshek \& Rao (2005)]{ntr05}
        Nestor, D. B., Turnshek, D. A., Rao, S. M., 2005, ApJ,
	628, 637

\bibitem[Noterdaeme et al. (2008)]{not08}%H2 survey
        Noterdaeme, P., Ledoux, C., Petitjean, P.,  
	Srianand, R.,   2008, A\&A, 481, 327

\bibitem[Peroux et al. (2003a)]{sub03}
        P\'eroux, C., Dessauges-Zavadsky, M., D'Odorico, S., Kim, T.-S.,
 	McMahon, R., 2003a, MNRAS, 345, 480

\bibitem[Peroux et al (2003b)]{cel03}
        Peroux, C.,  McMahon, R. G., Storrie-Lombardi, L. J., Irwin, M.,
        Hook, I. M., 2003b, MNRAS, 346, 1103

%\bibitem[Peroux et al (2007)]{per07}%Z evolution
%        Peroux, C., Dessauges-Zavadsky, M., D'Odorico, S., Kim, T.-S., 
%	McMahon, R. G., 2007, MNRAS, 382, 177

\bibitem[Peroux (2008)]{per08}
        Peroux, C., Meiring, J. D., Kulkarni, V. P., Khare, P., 
	Lauroesch, J. T., Vladilo, G., York, D. G., 2008, MNRAS, 386, 2209

\bibitem[Pettini et al. (1999)]{pet99}
	Pettini, M., Ellison, S. L., Steidel, C. C., Bowen, D. V.
	1999, ApJ, 510, 576

\bibitem[Pettini et al. (2000)]{pet00}
	Pettini, M., Ellison, S. L., Steidel, C. C., 
        Shapley, A. E., \& Bowen, D. V.
	2000, ApJ, 532, 65

\bibitem[Pettini et al. (2002)]{petno}
	Pettini, M., Ellison, S. L., Bergeron, J., Petitjean, P.,
	2002, A\&A, 391, 21

\bibitem[Prantzos and Boissier (2000)]{pb00}%zn-nhi
	Prantzos, N., Boissier, S., 2000, MNRAS, 315, 82

\bibitem[Prochaska and Wolfe (1997)]{pw97}
	Prochaska, J. X., \& Wolfe, A. M.
	1997, ApJ, 487, 73 % disks

\bibitem[Prochaska et al. (2001)]{ucsd1}
	Prochaska, J. X., Wolfe, A., Tytler, D., Burles, S., Cooke, J.,
	Gawiser, E., Kirkman, E., O'Meara, J., Storrie-Lombardi, L.,
	2001, ApJS, 137, 21.

\bibitem[Prochaska, Herbert-Fort \& Wolfe (2005)]{phw05}
        Prochaska, J. X., Herbert-Fort, S., \& Wolfe, A. M., 
	2005, ApJ, 635, 123

\bibitem[Prochaska et al. (2007a)]{jxp07}% echelle grb
        Prochaska et al., 2007a, ApJS, 168, 231

\bibitem[Prochaska et al. (2007b)]{pro07} % decade of keck
         Prochaska, J. X., Wolfe, A. M., Howk, J. C., Gawiser, E., 
	 Burles, S. M., Cooke, J., 2007b, ApJS, 171, 29

\bibitem[Prochter et al. (2006)]{proc06_mg2}
        Prochter, G. E., Prochaska, J. X., Burles, S. M.,
	2006a, ApJ, 639, 766

\bibitem[Prochter et al. (2006)]{proc06_grb}
        Prochter, G. E., Prochaska, J. X., Chen, H.-W., Bloom, J. S.,
	Dessauges-Zavadsky, M., Foley, R. J., Pettini, M.,
	Dupree, A. K., Guhathakurta, P., ApJ, 2006b, 648, L93

\bibitem[Rao and Turnshek (2000)]{rt00}
	Rao, S.M., \& Turnshek, D.A.
	2000, ApJS, 130, 1

\bibitem[Rao, Turnshek \& Nestor (2006)]{rtn06}
	Rao, S.M., Turnshek, D.A., Nestor, D. B., 2006, ApJ, 
	636, 610

\bibitem[Shapley et al (2003)]{alice03}
        Shapley, A., Steidel, C., Pettini, M., Adelberger, K.,  2003, 
	ApJ, 588, 65

\bibitem[Turnshek \& Rao (2002)]{tr02}
        Turnshek, D. A., Rao, S. M., 2002, ApJ, 572, L7

\bibitem[Vladilo et al. (2008)]{vpw08}
        Vladilo, G., Prochaska, J. X., Wolfe, A. M., 2008, A\&A, 478, 701

\bibitem[Vreeswijk et al (2007)]{time_var}
         Vreeswijk et al., 2007, A\&A, 468, 83

\bibitem[Weiner et al. (2008)]{wei08}
        Weiner, B. J., et al., 2008, ApJ, submitted, arXiv:0804.4686

%\bibitem[Wolfe, Gawiser \& Prochaska (2005)]{wgp05}
%        Wolfe, A., Gawiser, E., \& Prochaska, J. X., 2005, ARA\&A, 
%	43, 861

\bibitem[Zibetti et al. (2007)]{zib07}
	Zibetti, s., M\'enard, B., Nestor, D. B., Quider, A. M., Rao, 
	S. M., Turnshek, D. A., 2007, ApJ, 658, 161

\bibitem[Zych et al. (2008)]{bz08}
        Zych, B. J., Murphy, M. T.,  Hewett, P. C., Prochaska, J. X., MNRAS,
	2008, submitted.

\end{thebibliography}
\end{document}